\documentstyle[twocolumn,aps]{revtex}

\renewcommand{\abstractname}{ }
\makeatletter
\long\def\abstract#1{\def\@abstract{#1}}
\def\@abstract{}
\let\@oldmaketitle\@maketitle
\def\@maketitle{
 \@oldmaketitle
 \centerline{\bfseries\abstractname}
 \begin{quotation}\@abstract\end{quotation}
 \vskip-5mm}
\makeatother
\begin{document}
\draft
\preprint{}
\title{
Soft x-ray spectroscopy experiments on the near K-edge of B\\
in MB$_2$ (M=Mg, Al, Ta, and Nb)
}
\author{J. Nakamura, N. Yamada, and K. Kuroki}
\address{
Department of Applied Physics and Chemistry, 
The University of Electro-Communications,
Chofu-shi, Tokyo 182-8585, Japan
}
\author{T.A. Callcott}
\address{
Department of Physics, University of Tennessee,
Knoxville, TN 37996
}
\author{D.L. Ederer}
\address{
Department of Physics, Tulane University,
New Orleans, LA 70118
}
\author{J.D. Denlinger}
\address{
Advanced Light Source, Lawrence Berkeley National Laboratory,
Berkeley, CA 94720
}
\author{R.C.C. Perera}
\address{
Center for X-ray Optics, Lawrence Berkeley National Laboratory,
Berkeley, CA 94720
}
\date{Received \ \ \ \ \ \ \ \ \ \ \ \ \ \ \ \ \ \ }
%\maketitle
%\begin{abstract}
\abstract{\small
Soft X-ray absorption and emission measurements are performed for the K-
edge of B in MB$_2$ (M=Mg, Al, Ta and Nb).
Unique feature of MgB$_2$ with a high density of B 2$p_{xy}(\sigma)$-state
below and above the Fermi edge, which extends to 1 eV above the edge, 
is confirmed.
In contrast, the B 2$p$ density of states in AlB$_2$ and TaB$_2$, 
both of occupied and unoccupied states, decreased linearly towards 
the Fermi energy and showed a dip at the Fermi energy.
Furthermore, there is a broadening of the peaks with $p\sigma$-character in
XES and XAS of AlB$_2$, which is due to the increase of three dimensionality 
in the $p\sigma$-band in AlB$_2$.
The DOS of NbB$_2$ has a dip just below the Fermi energy.
The present results indicate that the large DOS of B-2$p\sigma$ states
near the Fermi energy are crucial for the superconductivity of MgB$_2$. 
\\
\\
71.20.-b, 74.25.Jb, 78.70.DM, 78.70.En
}
%\end{abstract}
\maketitle
%\pacs{71.20.-b, 74.25.Jb, 78.70.DM, 78.70.En}

\narrowtext
Since the discovery of superconductivity in MgB$_2$ with a transition 
temperature, $T_{\rm c}$, of 39 K by Nagamatsu {\it et al.},\cite{Nagamatsu} 
large number of researches from experimental\cite{Kurumaev,Callcott} and 
theoretical point\cite{Imada} of 
view have been performed, to explain the superconducting properties and mechanism 
of this new high-$T_{\rm c}$ superconductor.
The observed $T_{\rm c}$ of $\sim 40$K seems to exceed the upper bound of the 
transition temperature ($\sim 30$K) estimated for conventional BCS-type 
superconductors. Hence it is very important to clarify whether its 
superconducting mechanism is conventional or not.
In this context, Bud'ko {\it et al.} reported a boron isotope 
effect of $\alpha$=0.26, which suggests that phonons are playing 
an important role in the occurrence of superconductivity 
in this compound.\cite{Bud'ko}
Tunneling results indicate an $s$-wave nature of superconductivity.
\cite{Schmidt}
Theoretically, band calculations have suggested strong 
electron-phonon coupling.
\cite{Kortus,Belashchenko,Satta,An,Serrato,Suzuki,Kong}
All this evidence supports the conventional BCS-type 
superconductivity in MgB$_2$.
However, a detailed analysis of the specific heat and recent 
high resolution XPS experiment indicate that the superconducting gap must be 
anisotropic or two-band-like.\cite{Wang,Tsuda}

In order to clarify the mechanism of high $T_c$ superconductivity in 
MgB$_2$, it is important to investigate the difference in the electronic 
states between MgB$_2$ and other related compounds: MB$_2$(M=Al, Ta and Nb).
MgB$_2$ is a superconductor with $T_c=39$ K as mentioned, 
while AlB$_2$ has been reported to be a non-superconductor.\cite{Slusky}
TaB$_2$ has been reported as a superconductor of 9.5 K by 
Kaczorowski {\it et al.},\cite{Kaczorowski} while it has 
recently been reported to be a non-superconductor by 
Gasparov {\it et al.}\cite{Gasparov}
$T_c$ of NbB$_2$ is also controversial: $T_{\rm c}$=6 K by 
Cooper {\it et al.}\cite{Cooper}, 
5.2 K by Akimitsu {\it et al.}\cite{Akimitsu} and  
0.62 K by Leyarovskaya and Leyarovski\cite{Leyar}, 
while Gasparov reported it as a  non-superconductor.\cite{Gasparov}

In the present study, we present X-ray emission (XES) and absorption 
spectra (XAS) near the boron (B) K edge in MB$_2$ (M=Mg, Al, Ta and Nb).  
XAS was measured by both the total fluorescence yield (TFY) and the total 
electron yield (TEY) measurements at the same time.  
XES and XAS studies are powerful tools to 
probe the filled and empty electronic states of a specific orbital.
The reason we choose boron is because the band calculations 
for MgB$_2$ indicate that the bands near the Fermi energy 
are mainly composed from boron 2$p$ orbitals.
By measuring the dipole transition between 2$p$ states and 1$s$ core level
of boron, we can specifically probe the partial density of states (PDOS) of 
B 2$p$ states. 
Furthermore, XES and XAS by TFY are not surface sensitive in contrast 
to photo-electron spectroscopy.

The commercial specimens from Rare-Metallic Co. were used as 
samples of MB$_2$(M=Mg, Al, Ta and Nb).
The specimens were examined by powder X-ray diffraction (XRD) measurements.

XRD measurements showed single phase MgB$_2$ type pattern for all specimens.
The dc magnetizations were measured with a SQUID magnetometer in the 
temperature range from 1.8 to 100 K.  
The temperature dependencies of the susceptibility indicate that the 
superconducting transition temperature of about 38 K for MgB$_2$, 
and no superconducting transition for TaB$_2$, NbB$_2$ and AlB$_2$ above 1.8 K.  

The soft X-ray emission and absorption spectroscopies were performed at 
BL-8.0.1 of Advanced Light Source (ALS) in LBNL.
The resolutions of emission and absorption spectra are 0.3 and 0.1 eV, 
respectively.
In order to calibrate energy, XAS by TEY were also measured at the well 
calibrated beam line BL- 6.3.2 of the ALS.  
The XAS for all MB$_2$ compounds obtained by TEY shows a sharp peak at 
about 193.8 eV, which is attributed to boron oxides,\cite{Jia}  
while the XAS obtained by TFY shows no detectable peak at about 194 eV.
These results indicate that there is a small amount of boron oxide only on the 
surface, but not inside the bulk.
Here, we present the XAS obtained by TFY, so the presented results are 
free from the influence of boron oxides.

Figure 1(a) shows XES ($\circ$) and XAS ($\bullet$) of MgB$_2$.
The sharp decrease of XES and XAS at about 186.3 eV is attributed to the Fermi 
energy measured from 1$s$ core level.
The solid line in Fig. 1(b) is the boron PDOS obtained from a band 
structure calculation\cite{Satta}, 
where we have taken into account the effect of the instrumental resolution 
by gaussian broadening.
The intensities of experimental XES and XAS in Fig. 1(a) are scaled to 
the theoretical PDOS in the energy region, 
$E \le$ 182 eV for XES and 187 eV $\le E \le$ 191 eV for XAS.
The sum of the experimental XES and XAS are also plotted in Fig. 1(b).
It can be seen that the overall feature of both XES and XAS, 
including the existence of a large PDOS around the Fermi energy, 
are remarkably well reproduced by the band structure calculation,
enabling us to attribute each observed structure to $p\pi$ and/or $p\sigma$
states. Namely, the existence of peaks A and B, 
which is consistent with recent studies\cite{Kurumaev,Callcott}, 
are characteristic of bonding $p\sigma$ states.\cite{Satta,Serrato}
The region C in the energy range from 187 to 191 eV is attributed to 
the $p\pi$ states. A sharp peak D at about 192 eV in XAS is 
reported to be a resonance peak of $p\pi^*$ state\cite{Callcott}, 
and also corresponds to antibonding $p\sigma^*$ state predicted by a band calculation.  
Thus peak D contains both the $p\sigma^*$ and resonance
state of $p\pi^*$ states.

Figure 2(a) shows XES and XAS of AlB$_2$.
The intensity of XES is normalized so that the area intensity 
coincides with that for MgB$_2$ below 
$E_{\rm F}$, while the intensity of XAS is scaled so that the intensity 
in the  high energy region, $E \ge$ 198 eV, coincides with that for MgB$_2$.
In the high energy region, XAS shows no strong characteristic peaks.
A broad tail of XES below 183 eV is similar to that of MgB$_2$, 
but the value of $E_{\rm F}$ shifts to be 187.5 eV. 
The form of XES of AlB$_2$ is broad compared to that of MgB$_2$.
Figure 2(b) shows experimental PDOS derived from the sum of XES and XAS.
A dip is observed at about 188 eV near the Fermi energy,
indicating that the B 2$p$ PDOS around the Fermi energy 
is drastically reduced compared to that in MgB$_2$.
This is the major difference between MgB$_2$ and AlB$_2$.

This difference can be understood from 
results of the band calculation for AlB$_2$.\cite{Suzuki} 
Namely, there are several factors that make the boron 2$p$
PDOS around the Fermi level in AlB$_2$ much smaller than in MgB$_2$.
First of all, the bonding $\sigma$ bands, whose tops are located above the 
Fermi level in MgB$_2$, are fully filled in AlB$_2$.
Secondly, the Fermi level is located at a point where the top of the bonding 
and bottom of the antibonding $\pi$ bands touch with each other at the K point.
If the system were purely two-dimensional, this would be a point where the 
DOS vanishes linearly as a function of energy.
Although the $\pi$ band is three dimensional, the above two-dimensional 
property remains because the system is anisotropic.

The difference between MgB$_2$ and AlB$_2$ can qualitatively
be understood within a simple rigid band model, namely by
simply shifting the Fermi energy as mentioned above.  
To be more precise, there are some quantitative differences, 
whose origin seems to lie beyond a rigid band picture. 
Namely, in AlB$_2$, the intensity of XAS just above the dip is larger 
than that in MgB$_2$, while the intensity of peaks A and D is suppressed.  
Looking again into the band calculation results, 
these features may be attributed to the 
increase of three dimensionality in AlB$_2$.   

The XES and XAS of TaB$_2$ are similar to those for AlB$_2$ except for 
a shift in the Fermi energy [Fig. 3] up to 188.6 eV, owing
to a larger band filling compared with AlB$_2$.  
The B 2$p$ PDOS at the Fermi energy is similar to that for AlB$_2$, 
so if TaB$_2$ is indeed superconducting, the difference 
between these two compounds should lie elsewhere.  

Figure 4 shows XES and XAS of NbB$_2$.
The Fermi energy is almost the same as that of TaB$_2$, 
but a considerable amount of DOS below the Fermi energy is observed.
A dip is also observed at $\sim$187 eV and is lower than the Fermi energy.
The form of the XAS indicates a flat PDOS above 189 eV and shows a small peak D.
The PDOS around the Fermi energy is not so small.  
The character of the 
states near the Fermi energy cannot be identified from the present results.

To summarize, the most characteristic feature in MgB$_2$ as compared to 
other related materials is the large B 2$p$ PDOS around the Fermi level.
Since this is partially attributed to the existence of the $p\sigma$ 
bonding band at the Fermi level, 
one may be tempted to consider that the $p\sigma$  band plays a 
crucial role in the occurrence of superconductivity in MgB$_2$.\cite{Imada}
This is indeed probable, but is not necessarily the case because the 
$p\pi$ band filling is also different between MgB$_2$ and other materials
as mentioned above, which should result in a large difference 
in the shape of the $p\pi$ band Fermi surfaces.
Let us note that the shape of the Fermi surfaces can play 
an essential role in the occurrence of superconductivity.
For example, in those mechanisms that exploit nesting between the Fermi 
surfaces of bonding and antibonding 
$\pi$ bands,\cite{Kuroki,Furukawa,Yamaji} the shape of the Fermi 
surfaces (namely the $\pi$ band filling) is crucial.
We believe that further studies are necessary to clarify this point.

We express our thanks to Dr. H. Takenaka of NTT-AT, Dr. Y. Muramatsu of 
SPring8, Japan Atomic Energy Research Institute (JAERI) and Dr. E. Gullikson 
of CXRO-LBNL for their helpful support with spectroscopic measurements.  
We thank Professor S. Suzuki for sending ref.\cite{Suzuki} prior to publication.  
We also thank Professor K. Kohn of Waseda University for the support 
susceptibility measurement.
We would like to thank Professor K. Asai of University of 
Electro-Communications and Professor J. Akimitsu of Aoyama Gakuin University 
for useful discussions.  
This work was supported by Office of Basic Energy Science of the 
Department of Energy under contract Nos. DE-AC03-76SF00098 and W-7405-Eng-48 
and National Science Foundation (NSF) Grant No. DMR-9017996 to the University of 
Tennessee.
D.L.E. and T.A.C. also wish to acknowledge support from NSF Grant DMR-9801804.

\begin{figure}
\caption{
(a) The observed XES ($\circ$) and XAS ($\bullet$) spectra of MgB$_2$.
(b) The sum of XES and XAS ($\Box$) and the theoretical PDOS (solid line) 
derived from FLAPW method broadened with experimental resolution.
}\label{FIG1}
\end{figure}
\begin{figure}
\caption{
(a) The observed XES ($\circ$) and XAS ($\bullet$) spectra of AlB$_2$.
(b) The sum of XES and XAS ($\Box$).
}\label{FIG2}
\end{figure}
\begin{figure}
\caption{
The observed XES ($\circ$) and XAS ($\bullet$) spectra of TaB$_2$.
}\label{FIG3}
\end{figure}
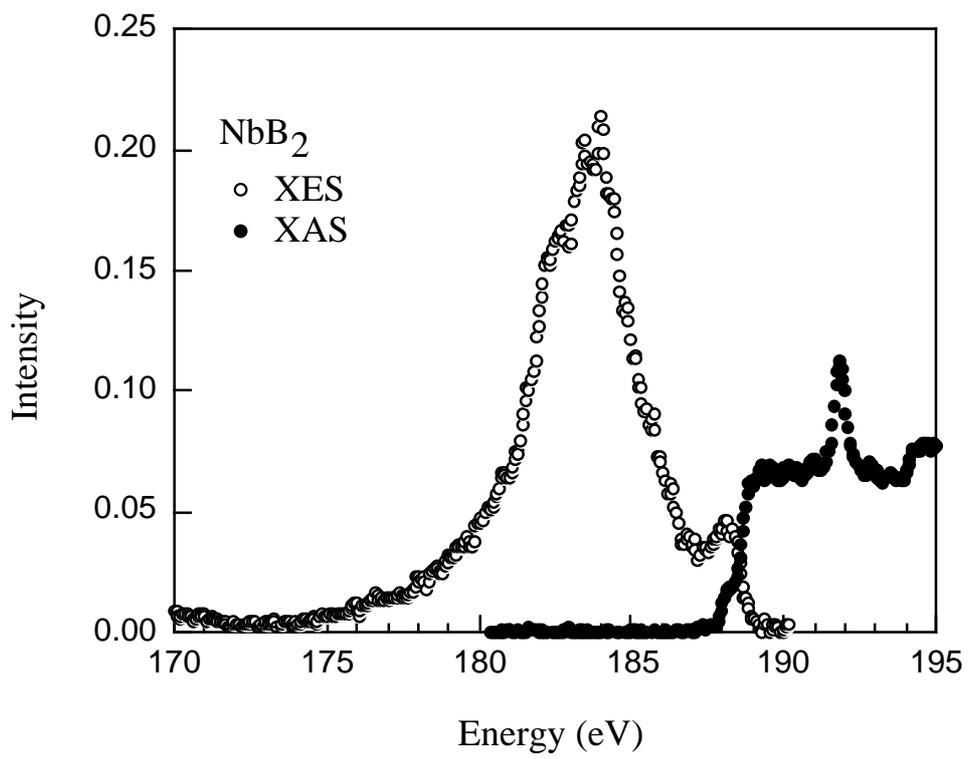
\begin{figure}
\caption{
The observed XES ($\circ$) and XAS ($\bullet$) spectra of NbB$_2$.
}\label{FIG4}
\end{figure}

\end{document}